\documentclass[letter, 10pt, conference]{ieeeconf}
\usepackage{comment}
\usepackage{article}
\IEEEoverridecommandlockouts
\title{\LARGE \bf
  Events! (Reactivity in urbiscript)
}

\author{Jean-Christophe Baillie \quad Akim Demaille \quad
  Quentin Hocquet \quad Matthieu Nottale \\
    Gostai S.A.S., 15, rue Jean-Baptiste Berlier, F-75013 Paris, France\\
    ~~~~{\small\url{http://www.gostai.com},
      \texttt{\var{first}.\var{last}@gostai.com}}%
}

\excludecomment{longVersion}

\begin{document}
\maketitle
\thispagestyle{empty}
\pagestyle{empty}

\begin{abstract}
  \usdk is a software platform for the development of portable robotic
  applications.  It features the Urbi UObject \Cxx middleware, to manage
  hardware drivers and/or possibly remote software components, and
  urbiscript, a domain specific programming language to orchestrate them.
  Reactivity is a key feature of Urbi SDK, embodied in events in \us.  This
  paper presents the support for events in \us.
\end{abstract}

Event-based programming is the ``native'' way in \us to program responses to
stimuli --- a common need in robotics.  It is typically used for
``background jobs that monitor some conditions''.  It is used to program the
human-robot interfaces (``do this when the head button is pushed''), the
detection exceptional situations (collision avoidance, battery level), the
tracking of objects of interest etc.  Events are also heavily used in the
implementation of Gostai Studio, a GUI for \urbi based on hierarchical
state machines \cite{gostai.10.studio}.

In following example ``whenever'' an object of interest (a ball) is visible,
the head of the robot is moved to track it.  The code relies on some of key
features of \us: UObjects (\lstinline|ball|, \lstinline|camera|,
\lstinline|headPitch|, \lstinline|headYaw|), concurrency (\lstinline|&|),
and event constructs (\lstinline|whenever|).

\begin{urbiunchecked}
whenever (ball.visible)
 {  headYaw.val   += camera.xfov * ball.x
  & headPitch.val += camera.yfov * ball.y };
\end{urbiunchecked}

\section{\urbi and \us}

\subsection{The \urbi Platform}
The \urbi platform \cite{baillie.05.iros}, including the \us programming
language, was designed to be \emph{simple and powerful}.  It targets a wide
spectrum of users, from children customizing their robots, to
researchers who want to focus on complex scientific problems in robotics
rather that on idiosyncrasies of some robot's specific \ac{api}.

\urbi relies on modularity to provide portability.  Components
specific to an architecture (sensors and actuators drivers, \ldots) are
implemented as \emph{UObjects},
plugged into the \urbi core.  UObjects can also wrap pure software components
that provide core services, say text-to-speech, object tracking,
localization, etc.  The \Cxx UObject architecture is actually a middleware:
components can be relocated transparently on remote computers instead of
running on the robot itself.  This is especially useful for low-end robots
whose CPU is too limited to run demanding software components (e.g., face
detection, speech recognition, \ldots).

The \urbi platform schedules the execution of these UObjects, routes the
events from producers to consumers, and so forth.  Its core is
written in \Cxx.  This choice was driven by the availability of compilers
for many different types of hardware architecture, and because many robot
SDK are in \C/\Cxx.  It also provides access to very low-level system
features (such as coroutines, see below), and allows us to program their
support if they lack, in assembler if needed.  Specific features of some
architectures are also easier to use from \C/\Cxx, such as the real-time
features of Xenomai.  Finally, some architecture \emph{require} \Cxx, such
the Aibo SDK.

While the sole \urbi core suffices in many situations, it proved useful to
provide a programming language to fine-tune this orchestration.

\subsection{The \us Programming Language}

There are already so many programming languages.  Why a new one?  Why not
extending an existing language, or relying on some library extensions?

\acp{dsl} are gaining audience because they make developers much more
productive than when using
the library-based approaches.  Programming robots requires a complete
rethinking of the execution model of traditional programming languages:
concurrency is the rule, not the exception, and event-driven programming is
the corner stone, not just a nice idiom.  These observations alone justify
the need for innovative programming languages.

There are already many well-established environments that provide these
features, and that can be used to program robots.  The world of synchronous
languages includes several adequate members, such as Lustre
\cite{halbwachs.91.ieee} or Esterel \cite{berry.92.scp}.  These systems
offer soundness and strong guarantees, but at a price: they are very
different from the programming languages developers are used to.  They are
adequate to develop real-time, life-critical systems, but they are too
demanding when developing the global behavior of a personal robot.  Some
general purpose programming languages have been extended also to offer
reactive programming: \C \cite{boussinot.91.spe}, Caml
\cite{mandel.05.ppdp}, Haskell \cite{peterson.99.icra}, etc.

Since the \urbi core is tightly bound to \Cxx, none of these languages
are adequate.  Binding with low-level languages (such as \Cxx)
is a domain in which scripting languages, such as Python \cite{rossum.95.tr}
or Lua \cite{ierusalimschy.96.spe}, excel.  It is not surprising that they
are acclaimed interfaces for practical robot programming environments such as
ROS \cite{quigley.09.icra}.

Yet, they do not provide native support for concurrency and reactivity, even
if there are existing extensions \cite{moura.04.jucs,tismer.00.tr}.  When
the \urbi project was started (circa 2003), the need for a new language,
tailored for programming robotic applications, was felt.

\medskip

To cope with the resistance to new languages, \us stays in the (syntactic)
spirit of some major programming languages: \Cxx, \Java, \Js etc.  As most
scripting languages, it is dynamically typed.

It is an \ac{ool}: values are \dfn{objects}.  Unlike most \acp{ool}, \us
is \emph{not} class-based.  In \dfn{class-based} \acp{ool} (such as \Cxx,
\Java, \Cs, Smalltalk, Python, Ruby, \ldots), \dfn{classes} are templates
(molds) that describe the behaviors and members of an object.  Classes are
\dfn{instantiated} to create a value; for instance the \lstinline|Point|
class serves as a template to create values such as
\lstinline|one = (1, 1)|.
The object \lstinline|one| holds the (dynamic) values while the class
captures the (static) behavior.  Inheritance in class-based languages is
between classes.

\us is \dfn{prototype-based}, like Self \cite{ungar.87.oopsla}, Lisaac
\cite{sonntag.02.tools}, Cecil \cite{chambers.04.cecil}, Io
\cite{dekorte.05.oopsla} and others.  In these \acp{ool}, there are no
classes.  Instantiation is replaced by \dfn{cloning}: an object serves as a
template for a fresh object, and inheritance relates objects\footnote{%
  Lines starting with a timestamp such as \lstinline|[00001451]| (1.1451s
  elapsed since the server was launched) are output by \urbi; the other
  lines were entered by the user.  The system answers with the value of the
  entered expressions, unless there is none (e.g., \lstinline|void|).  Due
  to space constraints, the system answers for functions (their definition)
  is not displayed in this paper.
}.

\begin{urbiscript}
// Create an empty object that derives from Object.
var one = Object.new();
[00000001] Object_0x109fce310
\end{urbiscript}

An object is composed of a list of \dfn{prototypes} (parent objects) and a
list of \dfn{slots}.  A slot maps an identifier to a value (an object).

\begin{urbiscript}
// one is a clone of Object.
one.protos;
[00000002] [Object]

// Add two slots, named x and y.
var one.x = 1;
[00000003] 1
var one.y = 1;
[00000004] 1
// The names of local slots (inherited slots
// are not reported).
one.localSlotNames;
[00000005] ["x", "y"]
\end{urbiscript}

\us is fully dynamic.  Objects can be extended at run-time: prototypes and
slots can be added or removed.  Functions are \dfn{first-class entities}:
they are ordinary values that can be assigned to variables, passed as
arguments to functions and so forth.  \us supports \dfn{closures}: functions
can capture references from their environment, then later use those
references to retrieve or set their content.  \us is a functional
programming language, functions are values that can be bound by slot like
any other object.

\begin{urbiscript}
one;
[00000006] Object_0x109fce310
// The function "asString" is used by the system
// to report values to the user.
function one.asString() { "(
one;
[00000007] (1, 1)
\end{urbiscript}

For a thorough presentation of \us, see \cite{gostai.10.usdk}.

\subsection{Concurrency}

Today, any computer runs many programs concurrently.  The Operating System
is in charge of providing each process with the illusion of a fresh machine
for it only, and of scheduling all these concurrent processes.  In robots,
jobs not only run concurrently, they heavily \emph{collaborate}.  Each job
is a part of a complex behavior that emerges from their
coordination/orchestration.

To support the development of concurrent programs, \us provides specific
control flow constructs.  In addition to the traditional sequential
composition with \samp{;}, \us provides the \samp{,} connector, which
launches the first statement in background, and immediately proceeds to
executing the next statements.  Scopes (statements enclosed in curly braces:
\lstinline|{ \var{s1}; \var{s2}, ...}|), are boundaries: a compound
statement ``ends'' when all its components did.  The following example
demonstrates these points.

\begin{urbiscript}
// "1s" means one second.  Launch two commands in
// background, using ",".  Execution flow exits the
// scope when they are done.
{ { sleep(2s); echo(2) }, { sleep(1s); echo(1) }, };
echo(3);
[00001451] *** 1
[00002447] *** 2
[00002447] *** 3
\end{urbiscript}

Other control flow constructs, such as loops, can be executed concurrently.
For instance, iterating over a collection comes in several flavors:
\lstinline|for| is sequential while \lstinline|for&| launches the iterations
concurrently (see below).

\vspace{-1em}
\begin{multicols}{2}
\begin{urbiscript}
for  (var i : [2,1,0]) {
  echo("
  sleep(i);
  echo("
};
echo("done");
[00125189] *** 2: start
[00127190] *** 2: done
[00127190] *** 1: start
[00128192] *** 1: done
[00128192] *** 0: start
[00128193] *** 0: done
[00128194] *** done
\end{urbiscript}
  \columnbreak
\begin{urbiscript}
for& (var i : [2,1,0]) {
  echo("
  sleep(i);
  echo("
};
echo("done");
[00105789] *** 2: start
[00105789] *** 1: start
[00105789] *** 0: start
[00105793] *** 0: done
[00106793] *** 1: done
[00107793] *** 2: done
[00107795] *** done
\end{urbiscript}
\end{multicols}
\vspace{-1em}

The standard library also provides functions that launch new tasks, running
in the background.  From an implementation point of view, \urbi relies on a
library of coroutines \cite{moura.04.tpls}, not system threads.  Job control
is provided by \emph{Tags}, whose description fall out of the scope of this
paper, see \cite{baillie.10.car}.

\begin{longVersion}
\begin{urbiscript}
// Instantiate a new Tag.
var t = Tag.new;
[00000010] Tag<tag_8>

// Run two commands concurrently, in background,
// under the control of the Tag t.
t : every (1s) echo("tick!"),
[00000019] *** tick!
sleep(0.5s);
t : every (1s) echo("tack!"),
[00000020] *** tack!
sleep(2s);
[00000021] *** tick!
[00000022] *** tack!
[00000023] *** tick!
[00000024] *** tack!

// Kill the jobs controlled by the Tag t.
t.stop;
sleep(2s);
// Nothing else runs.
\end{urbiscript}
\end{longVersion}

\section{Events}

\begin{longVersion}
  In order to facilitate the implementation of reaction to stimuli, \us
  features \dfn{Events}.
\end{longVersion}

\subsection{Basic Events}
Their use goes in three parts.  First, an event is needed, a derivative from
the \lstinline{Event} object.  This object will serve as an identifier for a
set of events, it can be used many times (or not at all).

\begin{urbiscript}
var e = Event.new;
[00007599] Event_0x104bcec50
\end{urbiscript}

Second, event handlers are needed.  They can catch any emission of an event,
or filter on the arity of the payload:

\begin{urbiscript}
at (e?)
  echo("e?");
at (e?())
  echo("e?()");
at (e?(var x))
  echo("e?(
at (e?(var x, var y))
  echo("e?(
\end{urbiscript}

Finally, we need to emit an event, possibly with a payload.

\begin{urbiscript}
e!;
[00000033] *** e?
[00000034] *** e?()

e!(12, "foo");
[00000035] *** e?
[00000036] *** e?(12, foo)

e!(12, "foo", 666);
[00000037] *** e?
\end{urbiscript}

The expression \lstinline|\var{e}!(\var{arg})| is syntactic sugar for
\lstinline|\var{e}.emit(\var{arg})|.

\subsection{Semantics}

The semantics is simple.  There is no guarantee on the order in which the
event handlers are run (or rather the order is an implementation detail that
is not enforced by our language definition).  The handling of events is
asynchronous by default, i.e., the control flow that emitted the event may
proceed before the event handlers are finished.

\begin{urbiscript}[firstnumber=1]
var f = Event.new;
[00000917] Event_0x107545d90

at (f?(var e)) { echo(e); sleep(0.5s); echo(e); };
f!("handler"); echo("top");
[00000919] *** handler
[00000928] *** top

sleep(1s); // Wait for the second echo.
[00001433] *** handler
\end{urbiscript}

\noindent
Event can be send synchronously to override this behavior.

\begin{urbiscript}
f.syncEmit("handler"); echo("top");
[00001929] *** handler
[00002436] *** handler
[00002436] *** top
\end{urbiscript}

Several handlers can run concurrently.

\begin{urbiscript}
f!("h1"); echo (1); f!("h2"); echo(2);
[00002437] *** h1
[00002442] *** 1
[00002442] *** h2
[00002448] *** 2
sleep(2s);
[00002942] *** h1
[00002954] *** h2
\end{urbiscript}

Event handlers may raise events; the system does not enforce laws that would
prevent endless constructs.  We believe that soundness properties such as
termination guarantees do not belong to the \urbi system, but to the
programmer.

\begin{urbiunchecked}[firstnumber=1]
var e = Event.new;
at (e?(var p)) { echo(p); e!(-p) };
e!(-1);
[00003919] *** -1
[00003925] *** 1
[00003931] *** -1
[00003936] *** 1
[00003945] *** -1
...
\end{urbiunchecked}

From the implementation point of view, a major constraint was the pressure
over the CPU.  Because in embedded systems (and in particular with low-cost
robots) the batteries must be saved to all cost, the implementation aims at
the lowest possible foot-print.  There is no active wait: in \urbi, if there
are no current computations but only event-constructs that monitor external
events (incoming network data, UObjects-generated events etc.), then the
system consumes no CPU at all.

It is on top of this event-handling layer that \us provides its supports for
monitoring arbitrary expressions, see \autoref{sec:exp}.

\subsection{Filtering}

Finally, \lstinline|at| clauses can filter on the payload.

\begin{urbiscript}[firstnumber=1]
var e = Event.new;
[00000002] Event_0xADDR
var x = 123;
[00000003] 123
at (e?(var x, var y) if x == y)
  echo("e?(
at (e?(x, var y))
  echo("e?(

e!(12, 34);

e!(12, 12);
[00000215] *** e?(12, 12) with 12 == 12

e!(x,  34);
[00000218] *** e?(123, 34)

e!(x,   x);
[00000221] *** e?(123, 123) with 123 == 123
[00000221] *** e?(123, 123)
\end{urbiscript}

\us supports pattern-matching, which can be used to filter the event
payload.  Special syntactic support is provided for tuples, lists, and
dictionaries (and their combinations).

\begin{urbiscript}[firstnumber=1]
var e = Event.new;
[00000206] Event_0x109324980

// Filter on lists.
at (e?([1, var y, "foo"]))
  echo("[1, 
e!(1, 2, "foo");
e!([1, 2, "foo"]);
[00000312] *** [1, 2, "foo"]
\end{urbiscript}

\begin{longVersion}
\begin{urbiscript}
// Filter on tuples.
at (e?((1, var y, "foo")))
  echo("(1, 
e!((1, 2, "bar"));
e!(1, 2, "foo");
e!((1, 2, "foo"));
[00000307] *** (1, 2, "foo")

// Filter on dictionaries.
at (e?(["one" => 1, "two" => var x]))
  echo("[\"one\" => 1, \"two\" => 
e!(["two" => 2]);
e!(["one" => 1, "two" => 2]);
[00000317] *** ["one" => 1, "two" => 2]

// Filter on combinations.
at (e?(["list"  => [1, var x],
        "tuple" => (1, var y)])
    if x == y)
  echo("list and tuple (1, 
e! (["list" => [1, 2], "tuple" => (1, 3)]);
e! (["list" => [1, 2], "tuple" => (1, 2)]);
[00000325] *** list and tuple (1, 2)
\end{urbiscript}
\end{longVersion}

\section{Expressions as Events}
\label{sec:exp}

\begin{longVersion}
  \subsection{Examples}
\end{longVersion}

The \lstinline|at| blocks can also be used to monitor arbitrary expressions:

\begin{urbiscript}
var x = 1;
[00000001] 1

// The absence of question mark indicates we're
// monitoring an expression, not an event.
at (x <= 0)
  echo("x is negative")
onleave
  echo("x isn't negative anymore");

x = -1;
[00000002] -1
[00000003] *** x is negative
x = -2;
[00000004] -2
x =  1;
[00000005] 1
[00000006] *** x isn't negative anymore
\end{urbiscript}

This is actually syntactic sugar on top of \lstinline|makeEvent(\var{exp})|.
It turns any Boolean expression into an event that triggers when the
expression evaluates to true, and stops when it becomes false again.

\begin{longVersion}
\begin{urbiscript}
// The previous example could be desugared this
// way:
at (makeEvent(x <= 0)?)
  echo("x is negative");
\end{urbiscript}
\end{longVersion}

\subsection{Durations}

Gostai developed some robotic behaviors using event constructs,
for instance to trigger some reaction when a ball is visible, or ceases to be.
Because the object detection is not perfect, the robot sometimes appeared to
behave erratically.  This is easy to solve using some hysteresis mechanism,
but the result is cluttered uses of event constructs.  The same happens when
implementing behaviors such as ``do this when the head of the robot is
patted for two seconds''.

To address this issue \us provides support to monitor conditions that are
sustained for some specified amount of time: the handler may require an
event to be sustained for a given amount of time before being ``accepted''
(\lstinline|at (\var{exp} ~ \var{duration})|).  The optional
\lstinline|onleave| clause is run when the condition is invalidated.

\begin{urbiscript}[firstnumber=1]
var x = 0;
[00001855] 0
at (0 < x ~ 1s)
  echo("
onleave
  echo("

t0 = time();
[00001862] 1.19469

// x is 1 for 1 second.
x = 1; sleep(1s);
[00001863] 1
[00002865] *** 1.0: at

// The event was triggered, it is not rerun.
sleep(1s);

// Reset.
x = 0;
[00003869] 0
[00003869] *** 2.0: onleave
// x is not positive long enough.
x = 1; sleep(0.9s); x = 0; sleep(0.5s);
[00003874] 1
[00004777] 0
// x is positive long enough.
x = 1; sleep(0.5s); x = 2; sleep(0.5s);
[00005281] 1
[00005783] 2
[00006283] *** 4.4: at
\end{urbiscript}

\subsection{Main Usage: Arbitrary Event Monitoring}

The main interest of this feature is to be able to monitor in an
event-driven way objects without requiring any special facility from
their 
part. Consider for instance a ``car'' object, that has a ``fuel'' slot
indicating the remaining fuel level. With a classical approach, if we want
to trigger an alert when we're running out of gas, the car would have to
provide an adequate event, and trigger it each time the fuel is altered and
is under a given threshold.

\begin{urbiscript}[firstnumber=1]
class Car
{
  // The urbiscript constructor.
  function init() {
    var this.fuel = 1;
    // We must manually create a specific
    // event to signal low fuel level.
    var this.lowFuel = Event.new;
  };

  // We must never update fuel directly, but use
  // this setter, otherwise we might end up not
  // sending the lowFuel event.
  function updateFuel(var v) {
    var previous = fuel;
    fuel = v;
    var threshold = 0.05;
    // We must emit the event only if we just
    // passed below the threshold.
    if (previous >= threshold && v < threshold)
      lowFuel!;
  };
};
[00000735] Car

var car = Car.new;
[00000802] Car_0x109f322e0

at (car.lowFuel?)
  echo("Warning, running out of gas!");
\end{urbiscript}

This method has several cons. It forces the \lstinline|Car| implementer to
write a lot of boiler plate code, and to anticipate all its user needs and
provide the adequate events. For instance here we cannot monitor other
values of the fuel level. A more generic interface would be to provide an
event that triggers each time the fuel level changes.

\begin{urbiscript}[firstnumber=1]
class Car
{
  function init() {
    var this.fuel = 1;
    // We must manually create a specific
    // event to signal low fuel level.
    var this.fuelChanged = Event.new;
  };

  // We must never update fuel directly, but use
  // this setter.
  function updateFuel(var v) {
    var previous = v;
    fuel = v;
    // For convenience, the event carries the
    // previous and current values as payload.
    fuelChanged!(previous, fuel);
  };
};
[00000735] Car
var car = Car.new;
[00000783] Car_0x102ed0990

// Like before, give a warning on low fuel level.
at (car.fuelChanged?(var prev, var cur)
    if 0.05 < prev && cur <= 0.05)
  echo("Warning, running out of gas!");
// Also, check we do not overfill the tank.
at (car.fuelChanged?(var prev, var cur)
    if prev < 0.95 && 0.95 <= cur)
  stopFillingGas();
\end{urbiscript}

This approach allows the user to monitor the fuel level for any
condition. However, the \lstinline|Car| implementer still has to provide and
trigger manually events for changing variables. The \lstinline|Car| user can
monitor the fuel level in any way, but at the expense of verbose and complex
\lstinline|at| constructs.

With the use of \lstinline|at| on arbitrary expressions, we can get rid of all these
problems: the \lstinline|Car| implementer simply uses the fuel slot
naturally by directly affecting it values. Since nothing special has to be
done, he cannot forget to provide any monitoring event. The user can monitor
any condition on the fuel slot, in a more concise way.

\begin{urbiscript}[firstnumber=1]
class Car
{
  function init() { var this.fuel = 0.5 }
  // No wrapper needed, directly update "fuel".
};
[00000735] Car
var car = Car.new;
[00000737] Car_0x102db0d80

// Car has a fuel level slot, a float.
car.fuel;
[00000000] 0.5

// Although Car wasn't designed to emit fuel
// events, we can react to its variations.
at (car.fuel < 0.05)
  echo("Warning, almost out of gas!");
at (0.95 < car.fuel)
  stopFillingGas();
\end{urbiscript}

\subsection{General Purpose Usage: Eased Flow Control}

Another interesting usage of this feature is simplified control flow in some
situation. We can for instance separate a stop condition from an algorithm
with an \lstinline|at|: the algorithm is as soon as the condition is
verified. We can then perform the algorithm without worrying about checking
our exit cases.

The following function is an example of synchronous flow control.  It
searches a value in a list, sorted (dichotomy) or not (linear search).

\begin{urbiscript}
function find(var list, var val, var sorted)
{
  var begin = 0;
  var end = list.size;

  // First, factor all our stop cases with ats.

  // When we find the element, return its index.
  at (list[begin] == val)
    return begin;
  // When we exhaust our search space, fail.
  at (begin == end)
    return -1;

  // Now only the traversals remain.
  if (sorted)
    // Perform a dichotomy on sorted vectors.
    loop
    {
      var middle = ((begin + end) / 2).floor;
      if (list[middle] < val)
        begin = middle + 1
      else
        end = middle
    }
  else
    // Linear search on non-sorted vectors.
    loop
      begin++;
};
\end{urbiscript}

\subsection{Implementation}

The main concern is efficiency.
A trivial implementation would be simple busy-looping:
checking at each instant (at the end of every scheduler cycle) whether
the condition is true, and triggering the event if needed. This would
work, but is of course highly time and space (and energy!) expensive.

We could check the condition only at a given frequency, but then we
could miss some events, if the condition becomes true and then false
again between two checks. Moreover, this would imply a potential delay
between the realization of the condition and the triggering of the
\lstinline|at|,
which can be acceptable for reading some robotic sensor, but not in
our previous vector-search algorithm. This would anyway not be
optimal, since we would probably end up doing too many checks.

Our implementation uses a ``push'' model rather than a ``poll'' one:
instead of checking the condition arbitrarily to see whether it
changed, it's the modification of any value that might alter the
expression that will trigger the reevaluation of the expression, and
potential triggering of the at block. This is optimal, since the
condition is reevaluated only if there's any chance it changed: if
nothing related to your \lstinline|at| happens, it won't consume any
CPU cycle.

To achieve this, when an \lstinline|at| is first declared, the condition is
evaluated, and any mutable data encountered during this evaluation is hooked
for modification. Then every time some of this data is altered, we
reevaluate the condition and trigger the \lstinline|at| block if needed.
When the condition is reevaluated, the hooked variables list is also flushed
and rebuilt since it might differ, if we take another branch of a
conditional statement for instance.

\begin{urbiunchecked}[firstnumber=1]
var global = false;
[00000619] false

function test(var foo)
  { if (global) foo.a == 1 else foo.a == foo.b };

var foo = Object.new;
[00000620] Object_0x1091939d0
var foo.a = 1;
[00000621] 1
var foo.b = 2;
[00000621] 2

at ({echo("evaluate"); test(foo)})
  echo("True!")
onleave
  echo("False!");
[00000001] *** evaluate

// Hooked variables here are: global, foo,
// foo.a and foo.b.

// Altering foo.b triggers an evaluation,
// but does not alter the result or the hooked
// variables list.
foo.b = 3;
[00000002] *** evaluate
[00000002] 3
// Altering global makes the condition true.
global = true;
[00000003] *** evaluate
[00000004] *** True!
[00000002] true

// The list of hooked variables is now: global,
// foo, foo.a.

// Altering foo.b does not trigger a reevaluation,
// since it no longer participates.
foo.b = 2;
[00000006] 2
\end{urbiunchecked}

\begin{longVersion}
\begin{urbiunchecked}
// Etc
foo.a = 2;
[00000005] *** evaluate
[00000006] *** False!
[00000006] 2
global = false;
[00000007] *** evaluate
[00000008] *** True!
[00000006] false
\end{urbiunchecked}
\end{longVersion}


\section{Future Work}

\subsection{Handlers Synchronicity}
As of today, \us allows the emission of an event to declare whether the
event handlers must be performed immediately (synchronous), or ``later''
(asynchronous).
\begin{longVersion}
  The expression \lstinline|e!(arg)| is actually syntactic sugar for
  \lstinline|e.emit(arg)|, an asynchronous emission.  In these conditions,
  the system provide no guarantee as to \emph{when} the handlers will the
  run.  Alternatively, \lstinline|e.syncEmit(arg)| guarantees that all the
  handler bodies will be run before the next statement is run.  It is very
  similar to a set of procedure calls.
\end{longVersion}
But experience shows that some patterns are better treated when it is on the
handler side that synchronicity is chosen.

\subsection{Congestion Control}

As reported earlier, the (asynchronous) execution of event handlers may
result in several instances of event handlers being run concurrently.
Depending on the application this might be a feature, or a nuisance.

There are several ways to process events whose handling is long.  Similarly
to what the clock-directed loops in \us already do, bodies could be
forbidden to overlap by waiting for the previous handler to finish, they
might need to be preempted by newer event emissions, we could event manage
some priority scheme, with features that would allow to drop obsolete
events.

\subsection{Urbi}

The \urbi platform, as a middleware, already provides support for emitting
events at the \Cxx level, independently of the \us language.  But the
reception part is ongoing work.

\section*{Conclusion}

We have presented event-driven programming in \us, the
programming language of choice to orchestrate UObjects (\urbi components).
Together
with the primitive support for concurrency, it is the corner stone for the
\urbi paradigm for the development of robotic applications.  Although it
bears resemblance to reactive programming languages, the \us language is
made to make experimentation easier by offering more opportunities for the
addition of feature \emph{a posteriori}.  Expression-based events are one
typical example: they allow to monitor events in a non-intrusive way,
without requiring that hooks were previously deployed in the monitored
objects.

\medskip
\paragraph{Acknowledgments}
The authors would like to thank the anonymous reviewers for their
suggestions, insight and helpful comments.

\bibliographystyle{plain}
\bibliography{share/bib/acronyms,%
  share/bib/comp.lang.cecil,%
  share/bib/comp.lang.io,%
  share/bib/comp.lang.lisaac,%
  share/bib/comp.lang.lua,%
  share/bib/comp.lang.python,%
  share/bib/comp.lang.reactive,%
  share/bib/comp.lang.robotics,%
  share/bib/comp.lang.self,%
  share/bib/comp.lang.urbi,%
  share/bib/comp.programming.threads,%
  share/bib/comp.robotics}
\end{document}